\documentstyle[aps,prd,preprint,floats,epsfig,tighten]{revtex}
\newcommand{\ie}{{\it i.e.}}
\newcommand{\ms}{$\overline{\mbox{MS}}$}
\newcommand{\ams}{\mbox{$\alpha_{\overline{\mbox{\tiny MS}}}$}}
\newcommand{\amst}{\mbox{${\widetilde{\alpha}_{\overline{\mbox{\tiny MS}}}}$}}
\newcommand{\av}{\mbox{$\alpha_{V}$}}

\newcommand{\nfz}{\mbox{$N_{F,V}^{ (0) }$}}

\begin{document}
\draft
\title{An Analytic Extension of the \ms\ Renormalization
Scheme\thanks{Work supported in part by the Department of Energy, contract
DE--AC03--76SF00515, Deutsche Forschungsgemeinschaft, Reference \# Me
1543/1-1, and the Swedish Natural Science Research Council,
contract F--PD 11264--301.}}
\author{Stanley J. Brodsky, Mandeep~S.~Gill,
Michael~Melles\thanks{Present address:
Department of Physics, University of Durham, Durham, United Kingdom}, and
Johan Rathsman\thanks{Present address: Department of Rad. Sciences,
Uppsala University, Uppsala, Sweden}}
\address{Stanford Linear Accelerator Center \\
Stanford University, Stanford, California 94309}
\preprint{\hfill SLAC--PUB--7157 (Rev--2)}
\maketitle
 \begin{abstract}
The conventional definition of the  running coupling $\ams(\mu)$ in quantum
chromodynamics is based on a solution to the renormalization group equations
which treats quarks as either completely massless at a renormalization scale
$\mu$ above their thresholds or infinitely massive at a scale below them. The
coupling is thus nonanalytic at these thresholds. In this paper we present an
analytic extension of $\ams(\mu)$ which incorporates the  finite-mass quark
threshold effects into the running of the coupling. This is achieved by using a
commensurate scale relation to connect $\ams(\mu)$ to the physical $\alpha_V$
scheme at specific scales, thus naturally including finite quark masses. The
analytic-extension inherits the exact analyticity of the \av\ scheme and
matches the conventional $\overline {MS}$ scheme far above and below mass
thresholds.  Furthermore just as in \av\ scheme, there is no renormalization
scale ambiguity, since the position of the physical mass thresholds is
unambiguous.
\end{abstract}

\section{Motivation}

The running coupling in quantum chromodynamics (QCD)
in the modified minimal subtraction ($\overline{\mbox{MS}}$)
scheme \cite{msbar} and other dimensional regularization schemes
is traditionally constructed by solving the renormalization group
equations using perturbative approximants to the $\beta$ function which
change discontinuously at the quark mass thresholds
\cite{soper,Marciano,Santamaria}.
This is equivalent to using effective
Lagrangians with a fixed number of massless fermions in each energy range
between the quark mass thresholds.
Thus in the $\overline{\mbox{MS}}$ scheme, the
$\beta(\mu)$ function depends on the number of ``massless'' quarks
$N_{F}(\mu)$ which is taken as a step function of the renormalization
scale $\mu$. Matching conditions at threshold require the equivalence
of one effective theory with $n$ massless flavors to another effective
theory with one massive and ($n-1$) massless quarks.
It should be noted that this does not prevent one from including quark
masses in the \ms\ scheme. However, the quark masses do not enter into
the $\beta$ function since the running of the coupling is mass
independent.

The one-loop matching conditions \cite{Weinberg,Hall,Schucker} in the
$\overline{\mbox{MS}}$ scheme require the coupling to be continuous if the
matching is done at the quark masses, although the derivative is discontinuous.
In two-loop matching
\cite{Bernreuther_Wetzel,L_R_V_NPB,Chetyrkin_Kniehl_Steinhauser}
the coupling itself becomes discontinuous if the matching is done at
the quark masses, but it can be rendered continuous by modifying the
$\overline{\mbox{MS}}$ scheme \cite{Marciano}.
Recently the three-loop matching conditions have been computed
\cite{Chetyrkin_Kniehl_Steinhauser}, which together with the four-loop
$\beta$-function \cite{Larin_Ritbergen_Vermaseren_PLB}, give the
possibility to evolve the $\overline{\mbox{MS}}$ coupling
to four loops with massless quarks. This gives a reduced
dependence on the matching scale, as shown in
\cite{Rodrigo_Pich_Santamaria}, but possibly a nonphysical threshold
dependence. However,
in such a treatment the derivatives of the coupling remain discontinuous.
The inevitable result of the matching in a dimensional regularization
scheme is that the running of the $\overline{\mbox{MS}}$ coupling in
the renormalization scale is  nonanalytic -- nondifferentiable or
even discontinuous -- as the quark mass thresholds are crossed. Thus
there is an intrinsic difficulty in expressing physical, smooth
observables as an expansion in the $\overline {\mbox{MS}}$ coupling.
It is clearly necessary to restore the finite quark mass effects in
their entirety in order to restore analyticity.

Aesthetically, it is unnatural to characterize physical theories in
terms of an artificially-constructed renormalization scheme such as
$\overline{\mbox{MS}}$; it is more physical to use an effective charge
as determined from experiment to define the fundamental coupling
\cite{blm}. For example, in analogy to quantum electrodynamics, one
could choose to define the QCD coupling as the coefficient
$\alpha_V(Q)$ in the static limit of the scattering potential between
two heavy quark-antiquark test charges:
\begin{equation}
\label{eq:avdef}
V(Q^2) = - 4 \pi C_F {\alpha_V(Q) \over Q^2}
\end{equation}
at the momentum transfer $q^2 = t = -Q^2$, where $C_F=
({N_C^2-1})/({2N_C}) ={4}/{3}$ is the Casimir operator for the
fundamental representation in $SU(N_C)$, (with $N_C=3$ for QCD).  Such
an effective charge automatically incorporates the quark mass threshold
effects in the running, and thus it has an analytic
$\beta$ function. The \av\ scheme is particularly well-suited to
summing the effects of gluon exchange at low momentum transfer, such
as in evaluating the final-state interaction corrections to heavy
quark production \cite{Kuhn}, or in evaluating the hard-scattering
matrix elements underlying exclusive processes \cite{Robertson}. A
physical effective charge has the additional advantage that the
Appelquist-Carazzone decoupling theorem \cite{App75} is automatically
incorporated.

In this paper we shall construct an analytic extension of the \ams\
scheme, which we call \amst, by connecting the coupling directly to the
analytic  and physically-defined \av\ scheme. The necessity for
an analytic coupling has been emphasized by Shirkov and his
collaborators\cite{shirkov}.
Our
definition allows one to use a scheme based on dimensional regularization,
but which also, in a simple way,
treats mass effects properly  between the mass thresholds. Thus,
instead of having the number of effective flavors ($N_F$) change discontinuously
at (or nearby) the quark threshold, we obtain an analytic $N_F(\mu)$
which is a continuous function of the renormalization scale $\mu$ and
the quark masses $m_i$. Thus the analytically-extended scheme inherits
the mass dependence of the physical scheme. In addition, the
renormalization scale $\mu$ that appears  in the analytically-extended
scheme \amst\ is directly related to the momentum transfer appearing in
the \av\ scheme and thus has a definite and simple physical
interpretation\footnote{A somewhat similar approach has been tried in
\cite{shirkov}, but using the unphysical MOM renormalization scheme to
implement the mass thresholds.}.

The essential advantage of the modified scheme \amst\ is that it
provides an analytic interpolation of conventional dimensional
regularization expressions by utilizing the mass dependence of the
physical \av\ scheme. In effect, quark thresholds are treated
analytically to all orders in $m^2/Q^2$; \ie, the evolution of our analytically
extended coupling in the intermediate regions reflects the actual mass
dependence of a physical effective charge and the analytic properties of
particle production in a physical process.
Just as in Abelian QED, the mass dependence of the effective potential
and the analytically-extended scheme \amst\ reflects the analyticity of the
physical thresholds for particle production in the
crossed channel.  Furthermore, the definiteness of the dependence in the quark
masses automatically constrains the renormalization scale.
Alternatively, one could connect \ams\ to another
physical charge such as $\alpha_R$ defined from $e^+e^-$ annihilation.

Our approach should be compared with the standard treatment of quark
mass threshold effects in the \ms\ scheme.
For fixed order in $\alpha_{\mbox{\scriptsize{s}}}$
the corrections due to finite quark
mass threshold effects which we are considering in this paper
have been calculated for the hadronic
width of the Z-boson and the $\tau$ lepton semihadronic decay rate
\cite{Chetyrkin,Hoang,Soper_Surguladze,L_R_V_NPB}. The calculations
have been made both exactly to order ${\alpha}_{\overline{\mbox{\tiny MS}}}^2$
and as expansions in terms of $m^2/Q^2$ and $Q^2/m^2$ for light and
heavy quarks respectively.
Note that in principle the determination  of the finite
mass threshold effects for physical observables in
dimensional regularization schemes would require
a complete all-orders analysis of
the higher-twist  mass corrections to the effective Lagrangian of the
theory.

There are a number of other reasons to construct an analytic extension
of the \ams\ scheme:
\begin{itemize}

\item The comparison of the values of the coupling
$\alpha_{\mbox{\scriptsize{s}}}$ as determined from different
experiments and at different momentum scales  is an essential
test of QCD (for a recent review of existing  measurements see
\cite{Burrows}). One source of error is neglect of quark
masses in the determination of $\alpha_{\mbox{\scriptsize{s}}}$ and
in the subsequent running of the coupling from the scale where it
has been determined to the conventional reference scale, the
$Z$-boson mass.

\item Lattice calculations for the J/$\Psi$ and $\Upsilon$
spectra  now provide the most  precise determination of
$\alpha_{\mbox{\scriptsize{s}}}$ at low momentum
scales\cite{lepage,davies,lattice,lattice_update}. It is
important to know how finite quark mass effects enter into the
running of this value of $\alpha_{\mbox{\scriptsize{s}}}$  to
lower and higher energy scales with as small an error as
possible.

\item
Finite mass threshold effects in supersymmetric
grand
unified theories are important when analyzing the running
and unification of couplings over very large ranges. It
has been discussed, for example, in refs. \cite{clavelli,bmp}.
However, the scale used in the running and for the threshold
effects has not been related to the physical scale which is
naturally obtained in our approach.

\item
It is natural to unify theories by matching physical couplings
and masses at the unification scale. This can be accomplished in
the \av\ scheme or equivalently \amst.

\end{itemize}

\section{Details of $\alpha_V$ }

In the case of the Abelian theory, the coupling $\alpha_V$ derived from the
heavy lepton potential is equivalent to using the effective charge defined from
the running of the photon propagator.   In the non-Abelian theory the gluon
propagator is not gauge invariant; one thus has to turn to a physical gauge
invariant observable such as the heavy quark potential.

The effective charge
${\alpha_V(Q)}$, defined as in Eq.~(\ref{eq:avdef})
can be calculated as a perturbation expansion in \ams,
\begin{eqnarray}\label{eq:av}
\alpha_V(Q) & = &
\alpha_{\overline{\mbox{\tiny MS}}}(\mu)+
v_{1,\overline{\mbox{\tiny MS}}}\left({Q \over \mu}\right)
{\alpha_{\overline{\mbox{\tiny MS}}}^2(\mu) \over \pi}+
v_{2,\overline{\mbox{\tiny MS}}}\left({Q \over \mu}\right)
{\alpha_{\overline{\mbox{\tiny MS}}}^3(\mu) \over \pi^2}+ \cdots
\end{eqnarray}
The first two nontrivial terms in the perturbative series
have been computed in the $\overline{\mbox{MS}}$ scheme
\cite{Susskind,Fischler,Appelquist_Dine_Muzinich,Feinberg,Billoire,Peter}.
A comprehensive analysis of \av\ to order $\alpha^3$ has recently been given by
M. Peter \cite{Peter}:
\begin{eqnarray}
v_{1,\overline{\mbox{\tiny MS}}}(\mu=Q)
& = & -\frac{2}{3}N_C+\frac{5}{6}\beta_0 \; = \; 2.583-0.278N_F
\nonumber \\
v_{2,\overline{\mbox{\tiny MS}}}(\mu=Q) & = &
\left(\frac{133}{144}+\frac{24\pi^2-\pi^4}{64}-
\frac{11}{4}\zeta_3\right)N_C^2-
\left(\frac{385}{192}-\frac{11}{4}\zeta_3\right)C_F N_C+
\nonumber \\ &&
\frac{5}{6}\beta_1+
\left[\left(\frac{35}{32}-\frac{3}{2}\zeta_3\right)C_F+
\left(-\frac{217}{144}+\frac{7}{4}\zeta_3\right)N_C\right]\beta_0+
\frac{25}{36}\beta_0^2
\nonumber \\ & = &
39.650-4.147N_F+0.0772N_F^2
\end{eqnarray}
where $\beta_0$ and $\beta_1$ are the first two universal coefficients
in the $\beta$-function.
It is also known that the next coefficient in the expansion is
non-analytic in $\alpha=0$,
since it contains a $\ln(\alpha)$ term
\cite{Appelquist_Dine_Muzinich}.
This non-analyticity
does not originate from fermionic corrections to the heavy quark
potential. This can be seen by adopting a physical gauge. In such
gauges we would reproduce the same analyticity structure for fermionic
corrections in QCD as we find in QED. In QED, however, we have no
problem with analyticity at any order in perturbation theory.
Note that the total derivative
of $\alpha_V$ with respect to the renormalization scale $\mu$
of the \ams\ scheme is zero, since $\alpha_V$ is a physical observable.

The scale ($Q$) dependence of the effective charge defines the
equivalent of the Gell-Mann Low $\psi$-function for the effective
charge in QED \cite{GellMann_Low}. In the case of $\alpha_V$, $Q$ is the
momentum
transfer in the heavy quark potential, and the
$\psi$-function is given by,
\begin{equation}\label{eq:avrge}
\frac{d \alpha_V(Q)}{d \ln Q}= -\psi^{(0)}\frac{\alpha_V^2}{\pi}-
\psi^{(1)}\frac{\alpha_V^3}{\pi^2}-\psi_{V}^{(2)}\frac{\alpha_V^4}{\pi^3}-
\psi_{V}^{(3)}\frac{\alpha_V^5}{\pi^4}-\cdots \; \; .
\end{equation}
The first two terms in this series coincide with the universal and
well-known first two terms in the Callan-Symanzik $\beta$-function
\cite{Callan_Symanzik}, i.e. $\psi^{(0)}=\beta_0$
and $\psi^{(1)}=\beta_1$, but the higher order terms,
$\psi_{V}^{(2)}$ etc.,
depends on the observable under
study\footnote{Some authors denote the coefficient $\psi_{V}^{(2)}$ by
$\hat{\beta}_{2,V}$
instead. The convention used here is to emphasize the difference
between the dependence on the physical scale $Q$ and the unphysical
renormalization scale $\mu$.}.

For completeness we give the coefficients in the QCD $\psi$
function with the normalization used above,
\begin{eqnarray*}
\psi^{(0)} & = & \frac{11}{6}N_C-\frac{1}{3}N_F =  5.500-0.333N_F
\nonumber \\
\psi^{(1)} & = & \frac{17}{12}N_C^2-\frac{5}{12}N_CN_F-\frac{1}{4}C_FN_F
= 12.750 -1.583N_F
\nonumber \\
\psi_{V}^{(2)} & = &
\beta_{2,\overline{\mbox{\tiny MS}}}
- \psi^{(1)} v_{1,\overline{\mbox{\tiny MS}}}
- \psi^{(0)} v_{1,\overline{\mbox{\tiny MS}}}^2
+ \psi^{(0)} v_{2,\overline{\mbox{\tiny MS}}}
\nonumber \\ & = &
\left(\frac{103}{48}+\frac{11(24\pi^2-\pi^4)}{384}+
      \frac{121}{144}\zeta_3\right)N_C^3+
\nonumber \\ &&
\left[ \left(-\frac{445}{576}-\frac{24\pi^2-\pi^4}{192}
             -\frac{11}{9}\zeta_3 \right)N_C^2+
 \right.
\nonumber \\ &&
 \left.
       \left(-\frac{343}{288}+\frac{11}{12}\zeta_3 \right)N_CC_F +
       \frac{1}{32}C_F^2
\right]N_F+
\nonumber \\ &&
\left[ \left(\frac{1}{576}+\frac{7}{36}\zeta_3 \right)N_C+
       \left(\frac{23}{144}-\frac{1}{6}\zeta_3 \right)C_F
\right]N_F^2
\nonumber \\ & = &
193.074-27.014N_F+0.652N_F^2
\end{eqnarray*}
The results given above in the \ams\ scheme have been obtained using
massless QCD.  The effects of non-vanishing quark masses can be taken
into account by using a $Q^2$-dependent $N_F$ which will be derived
from the one-loop massive vacuum polarization function in the next
section.

\section{Calculation of the Running Coupling to One-Loop Order}

Our approach in this paper is as follows: the $\av(Q)$ scheme
automatically includes the effects of finite quark masses in the same
manner that lepton masses appear in Abelian QED.  We can then relate
the \ms\ scheme to the \av\ scheme through a commensurate scale
relation \cite{csr}, which is effectively a scale transformation
between the two schemes.  The analytic dependence of \av\ is then
transferred to the analytically-extended \amst\ scheme. The usual
massless expressions are recovered far above or far below any
individual quark mass threshold.

\subsection{Calculation of the Mass Dependence for the Running Coupling}

The coupling $\av(Q)$, which is derived from heavy quark scattering, is
closely related to the renormalization of the gluon propagator. In
physical gauges with $Z_1=Z_2$ the coupling renormalization is due
purely to self-energy insertions in the propagator.\footnote{Strictly
speaking, this is only true up to  one-loop in QCD and two-loops in QED.
At higher orders new types of diagrams appear in the potential which
cannot be described as simple self-energy insertions in the propagator.
In QCD such a diagram is the so called ``H-graph" \cite{Giles_McLerran}
and in QED the light-by-light scattering diagram has the same effect.
In the QED case, the light-by-light scattering graphs have an anomalous
dependence on the external charges and a cut structure corresponding to
particle production.
In addition, we
note that the non-analytic contributions to \av\ in higher orders in QCD arise
from corrections to the ``H-graph". Therefore it could be argued that these
types of diagrams should be excluded when defining the V-scheme in QCD and QED.}

For the purposes of this paper it will be sufficient to restrict our
analysis to one-loop order\footnote{We expect the main effects from
including the quark masses at the one-loop level as this is
the leading term in the $\psi$-function. However, at small scales
the higher order terms will become important, especially since
the relative importance of the $N_F$ term is larger for $\psi^{(1)}$
than for $\psi^{(0)}$. A study at the two-loop level requires the
massive two-loop diagrams which is work in progress \cite{bgmr2}.},
i.e. $\psi^{(0)}$.

The physical running coupling in the \av\ scheme, normalized at an
arbitrary momentum transfer scale $Q_0$, may be represented as
\begin{equation}
\label{running}
 \av(Q) \equiv \frac{\alpha_V(Q_0)}{1 -
\tilde{\Pi}\left(Q,Q_0,\alpha_V(Q_0)\right)} .
\end{equation}
The vacuum polarization function $\tilde{\Pi}$ may be computed from the
perturbative expansion of the renormalized propagator between heavy
quarks. The coupling is then
\begin{equation}
 \av(Q) =\alpha_V(Q_0) \left[ 1 + \tilde{\Pi} +
\tilde{\Pi}  ^2 +  \tilde{\Pi}  ^3 + \ .\ .\ . \right] ,
\end{equation}
where we have used the shorthand $ \tilde{\Pi}\equiv
\tilde{\Pi}(Q,Q_0,\alpha_V(Q_0)) $ for the renormalized sum of all
one-particle irreducible 1PI diagrams for the gluon self-energy. Since
the coupling has the value $\alpha_V(Q_0) \equiv \alpha_0 $ at the
physical renormalization point $Q=Q_0$, the self-energy obeys the
boundary condition $ \tilde{\Pi}(Q_0,Q_0,\alpha_0) = 0$.
\vspace{.5cm}

\begin{figure}[htbp]
\begin{center}
\leavevmode
\epsfbox{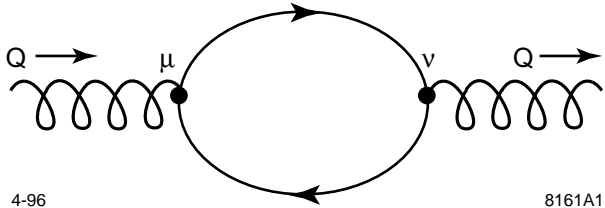}
\end{center}
\caption[*]{Single insertion of massive quark-antiquark loop into
a gluon propagator, giving the quark part of the one-loop gluon vacuum
polarization.}
\label{fig:onepidiag}
\end{figure}

\begin{figure}[htb]
\begin{center}
\mbox{\epsfig{figure=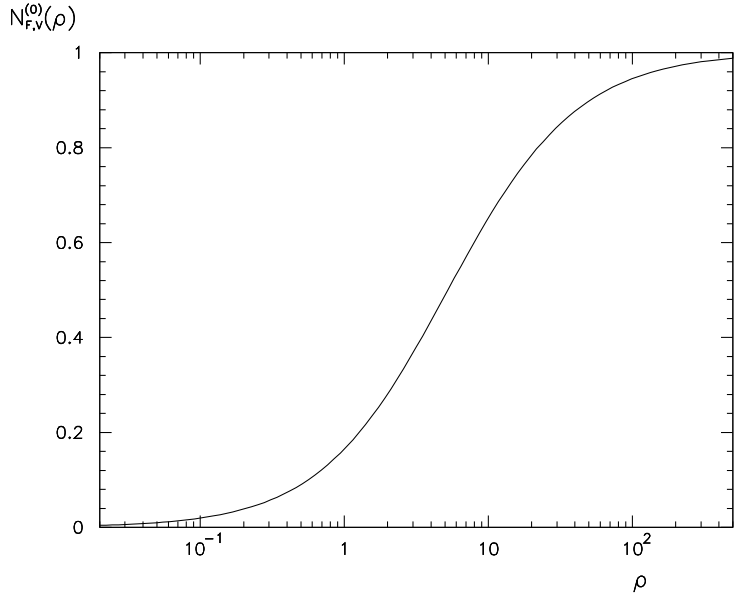,width=10cm}}
\end{center}
\caption[*]{The curve shows the contribution to the continuous
$\nfz$ for just one quark as a function of $\rho=Q^2/m^2$ where $m$ is
the mass of the quark. $\nfz$ is found by using the massive quark part
of the one-loop gluon propagator instead of using the theta function
thresholds conventionally used in dimensional regularization schemes.}
\label{fig:pnfo}
\end{figure}

We begin by considering the integral representation of the quark part
of the one-loop gluon vacuum
polarization diagram (see Fig.~\ref{fig:onepidiag}):
\begin{eqnarray*}
\tilde{\Pi}^{(0)}_{\mbox{\scriptsize{q}}}(Q,Q_0,\alpha_0) =
T_F \sum_{i=1}^{n} \frac{\alpha_0}{3 \pi} \left(
\int_{0}^{1} 6 z (1-z) \ln{\left(1+z (1-z) \rho_i(Q)\right)} - \right.
\\
\left. \int_{0}^{1} 6 z (1-z)
\ln{\left(1+z (1-z) \rho_i(Q_0)\right)} \right)
\end{eqnarray*}
where $\rho = Q^2/m^2$, $T_F={1 \over 2}$,
the superscript (0) indicates the one-loop order,
the subscript q indicates the quark-part
and the sum runs over all quarks ($n$).
Thus the quark component of the one-loop $\psi$-function is:
\begin{equation}
\psi_{V,\mbox{\scriptsize{q}}}^{(0)}(Q)
= -{\nfz \over 3}
= -\left[ \frac{\pi}{\av^2}
   \frac{d \av}{d \ln{Q}}\right]^{(0)}_{\mbox{\scriptsize{q}}}
=  -\frac{\pi}{\alpha_0}
\frac{d \, \tilde{\Pi}^{(0)}_{\mbox{\scriptsize{q}}}(Q,Q_0,\alpha_0)}
{d \ln{Q}}.
\end{equation}
This gives\footnote{This result was first obtained by Georgi and
Politzer \cite{Georgi_Politzer} in the MOM scheme and was applied to
general gauge theories by Ross\cite{Ross}. } the contribution to $N_F$
from quark flavor
$i$,
\begin{equation}
\nfz( \rho_i)= 6 \int_{0}^{1}
 \frac{z^{2}(1-z)^{2}\rho_i dz}{1+z(1-z)\rho_i}=
 1- \frac{6}{\rho_i} + \frac{24}{\rho_i^{3/2}\sqrt{4+\rho_i}}
         \tanh^{-1}{\sqrt{\frac{\rho_i}{\rho_i+4}}},
\end{equation}
which is displayed in Fig.~\ref{fig:pnfo} as a function of
$\rho$. Thus, by keeping the explicit quark mass dependence,
$N_F$ becomes an analytic function of the scale $Q$.

In fact, the approximate form:
\begin{equation}
\nfz(\rho_i) \cong \left(1 + {5 \over \rho_i} \right) ^{-1}
\label{eq:nfzappr}
\end{equation}
gives an accurate approximation to the exact form to within
a percent over the entire range of the momentum
transfer\footnote{This
approximate form can be obtained from using a rigorous
double asymptotic series approach, knowing the behavior of the function
at the low and high momentum transfer. \cite{oleg}}.

\begin{figure}[htb]
\begin{center}
\mbox{\epsfig{figure=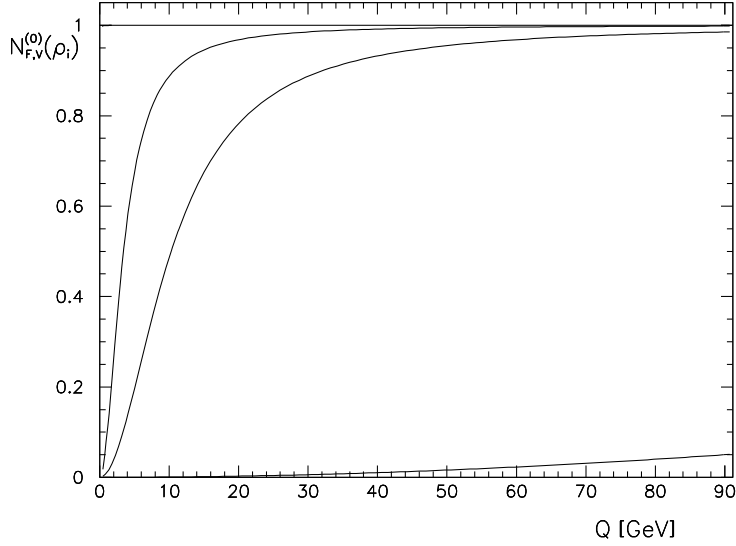,width=10cm}}
\end{center}
\caption[*]{Continuous \nfz\ for various quarks, lightest to heaviest goes
top to bottom
(d, c, b, t as one proceed downwards; the u and s plots are virtually
identical at this scale to the d).
Q runs from 1 to $M_Z$ GeV (for reference, the quark masses (in GeV) used are:
$m_u=.004$; $m_d=.008$; $m_s=.200$; $m_c=1.5$; $m_b=4.5$; $m_t=175$.)}
\label{fig:nfsep}
\end{figure}

The one-loop analytic $N_{F,V}$ is shown in Fig.~\ref{fig:nfsep} for
various quark flavors (for reference,
the quark masses (in GeV) we used are:
$m_u=.004$; $m_d=.008$; $m_s=.200$; $m_c=1.5$; $m_b=4.5$; $m_t=175$).

We may now substitute the $N_{F,V}$ into the one-loop QCD $\psi$ function
coefficient:
\begin{eqnarray*}
\psi_{V}^{(0)}(Q)= \frac{11}{2} - \frac{1}{3}  \nfz(\rho_i)
\end{eqnarray*}
and thence into the QCD one-loop renormalization group equation
for the coupling constant:
\begin{equation}
{ {d \av} \over d \ln{Q} } =
-  \psi_{V}^{(0)} { \av^2 \over \pi}
\end{equation}
We may then solve this renormalization group equation to yield an
expression for \av\  which is
analytic at mass thresholds. Note that the mass-dependence of the
$\psi$ function applies specifically to the \av\ scheme\footnote{Given
$\alpha_V(Q_0)$ one can obtain the coupling at other scales including
the mass dependence by numerical iteration such as the fourth-order
Runge-Kutta algorithm.}.

\subsection{Commensurate scale relation between \av\ and \ams\ }
We now relate the mass dependence of the \av\ scheme to the
$\overline{\mbox{MS}}$ scheme using the commensurate scale relation
\cite{blm,csr} between the  two schemes.  We use the NNLO results of
Peter\cite{Peter}. The first step is to invert
Eq.~(\ref{eq:av}) to obtain $\alpha_{\overline{\mbox{\tiny MS}}}$ as an
expansion in $\alpha_V$,
\begin{eqnarray}\label{eq:avinv}
\alpha_{\overline{\mbox{\tiny MS}}}(Q)
 & = &
\alpha_V(M)
+ m_{1,V}\left({Q \over M}\right)
{\alpha_V^2(M) \over \pi}+
m_{2,V}\left({Q \over M}\right)
{\alpha_V^3(M) \over \pi^2}+ \cdots \; .
\end{eqnarray}
The needed commensurate scale relation is obtained by fixing the scales
$M$ in Eq.~(\ref{eq:avinv}) such that the $\psi^{(0)}$ and $\psi^{(1)}$
dependent parts of the coefficients $m_{1,V}$ and $m_{2,V}$ are
absorbed into the running of the coupling $\alpha_V(M)$. This insures
that all vacuum polarization dependence is summed into the heavy quark
potential. Application of this procedure in next-to-next-to leading
order (NNLO), using the multi-scale approach \cite{csr}, gives the
following scale-fixed  relation between $\alpha_V$  and the
conventional $\overline{\mbox{MS}}$,
\begin{eqnarray} \label{eq:csrmsofv}
\alpha_{\overline{\mbox{\tiny MS}}}(Q)
& = & \alpha_V(Q^{*})
+ \frac{2}{3}N_C{\alpha_V^2(Q^{**}) \over \pi}
\nonumber \\ &&
+\left\{
-\left(\frac{5}{144}+\frac{24\pi^2-\pi^4}{64}-
       \frac{11}{4}\zeta_3\right)N_C^2
+\left(\frac{385}{192}-\frac{11}{4}\zeta_3\right)C_F N_C\right\}
 {\alpha_V^3(Q^{***}) \over \pi^2}
\nonumber \\ & = &
\alpha_V(Q^{*})
+ 2{\alpha_V^2(Q^{**}) \over \pi}
+ 4.625 {\alpha_V^3(Q^{***}) \over \pi^2} ,
\end{eqnarray}
above or below the quark mass threshold\footnote{Note that the
NNLO results depend
crucially on whether or not the ``H-graph" is included in the definition of
the heavy quark potential since it is the  unique source of the $\pi^4 N_C^2$
terms in the NNLO coefficient. We thank M. Peter for communications on this
point.}. The coefficients in the
perturbation expansion have their conformal values, i.e. the same
coefficients would occur even if the theory had been conformally
invariant with $\psi^{(0)}=0$ and thus do not contain the diverging
$(\psi^{(0)}\alpha_{\mbox{\scriptsize{s}}})^n n!$  growth
characteristic of an infrared renormalon \cite{Kataev}. The next-to
leading order (NLO) coefficient $\frac{2}{3}N_C$ is a feature of the
non-Abelian couplings of QCD and is not present in QED. The
commensurate scales $Q^*$ and $Q^{**}$ are given by
\begin{eqnarray}
Q^* & = & Q\exp\left[\frac{5}{6}\right] = 2.300 Q
\\
Q^{**} & = & Q\exp\left[
\left(\frac{105}{128}-\frac{9}{8}\zeta_3\right)\frac{C_F}{N_C}
+\left(\frac{103}{192}+\frac{21}{16}\zeta_3\right)\right] = 6.539 Q
\end{eqnarray}
whereas to this order $Q^{***}$ is not constrained.
However, a first approximation is
obtained by setting $Q^{***}=Q^{**}$.
Also note that $Q^*$ is unchanged when going from NLO to NNLO.
The scale $Q^*$ arises because of the
convention used in defining the modified minimal subtraction
scheme. Comparing the scales $Q$ and $Q^*$ we find that
the scale in the $\overline {\mbox{MS}}$ scheme ($Q$) is a factor
$\sim 0.4$ smaller than the physical scale ($Q^*$).

Alternatively, one can write the relation between
$\alpha_{\overline{\mbox{\tiny MS}}}$ and  $\alpha_V$ as a single-scale
commensurate scale relation \cite{Kataev}.  In this procedure $Q^* = Q^{**}$
where
\begin{eqnarray}
Q^* & = & Q\exp\left[\frac{5}{6}  +
    \left[\left(\frac{35}{32}-\frac{3}{2}\zeta_3\right)C_F -
        \left(\frac {19}{48} -\frac{7}{4}\zeta_3\right)N_C\right]
               \frac{\alpha_V}{\pi}
             \right]
\end{eqnarray}
The conformal coefficients are the same in the two
procedures\footnote{Both the multiple and single-scale
setting methods generate a term
proportional to $C_F N_C$ in the NNLO conformal coefficient. The origin of
this term, which has the same color factor as an iteration of the potential,
is not clear and should be further investigated.}.
However, the single-scale form has the advantage that the non-Abelian
perturbation theory matches in a simple way the corresponding Abelian
perturbation theory in the limit
$N_C
\to 0$ with
$C_F\alpha_s$ and
$N_F/C_F$ fixed \cite{Brodsky-Huet}.  For
$N_C= 3$ we have
$\ln(Q^*/Q) =   5/6 + 4.178 \alpha_V/\pi.$

\subsection{Definition of the Analytic \amst\ }

We now adopt the commensurate scale relation with the effective charge of the
effective potential as a definition of the  extended scheme \amst:
\begin{equation}
\widetilde {\alpha}_{\overline{\mbox{\tiny MS}}}(Q)
=  \alpha_V(Q^*) + \frac{2N_C}{3} {\alpha_V^2(Q^{**})\over\pi} +
\cdots ,
\label{alpmsbar2}
\end {equation}
for all scales $Q$. Eq.~(\ref{alpmsbar2}) not only provides an analytic
extension of dimensionally regulated schemes, but it also ties down the
renormalization scale to the physical masses of the quarks as they
enter into the vacuum polarization contributions to $\alpha_V$. There
is thus no scale ambiguity in perturbative expansions in \av\ or \amst.

Taking the logarithmic derivative of the commensurate scale relation
given by Eq.~(\ref{alpmsbar2})
with respect to $\ln Q$ we can define the $\psi$-function for the
\amst\ scheme as follows,
\begin{equation}
\widetilde {\psi}_{\overline{\mbox{\tiny MS}}}(Q)
\equiv \psi_V(Q^*) + 2 \frac{2N_C}{3} {\alpha_V(Q^{**})\over\pi}
\psi_V(Q^{**}).
\end{equation}
To lowest order this gives
$\widetilde {\psi}_{\overline{\mbox{\tiny MS}}}^{ (0) }(Q)
= \psi_V^{ (0) }(Q^*)$,
which in turn gives the following relation between
$\widetilde {N}_{F,\overline{\mbox{\tiny MS}}}^{(0)}$
and $N_{F,V}^{ (0) }$,
\begin{equation}
\widetilde {N}_{F,\overline{\mbox{\tiny MS}}}^{(0)}(Q)
=N_{F,V}^{ (0) }(Q^*),
\end{equation}
where to lowest order, $Q^* = \exp(5/6) Q$.

We can also use the approximate form given
by Eq.~(\ref{eq:nfzappr}) to write
\begin{equation}
\widetilde {N}_{F,\overline{\mbox{\tiny MS}}}^{(0)}(\rho_i)
\cong \left(1 + {5  \over \rho_i{\exp({5\over 3})}} \right)^{-1}
\cong \left( 1 + {1  \over {\rho_i}} \right)^{-1}.
\end{equation}
In other words the contribution from one flavor is $\simeq 0.5$ when
the scale $Q$ equals the quark mass $m_i$. Thus the standard procedure
of matching $\alpha_{\overline{\mbox{\tiny MS}}}(\mu)$ at the quark
masses is a zeroth order approximation to the continuous $N_F$.

\begin{figure}[htb]
\begin{center}
\mbox{\epsfig{figure=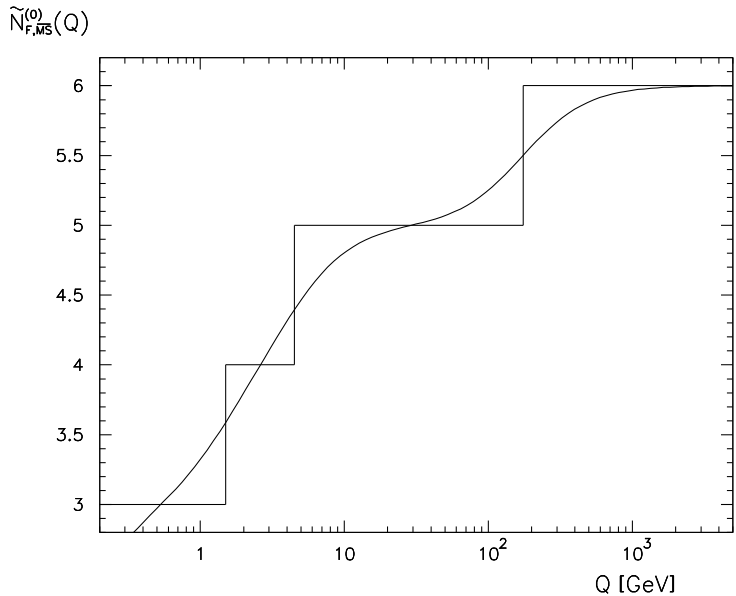,width=10cm}}
\end{center}
\caption[*]{The continuous
$\widetilde {N}_{F,\overline{\mbox{\tiny MS}}}^{(0)}$ in the analytic
extension of the $\overline{\mbox{MS}}$ scheme as a
function of the physical scale $Q$. (For reference the
continuous $N_F$ is also compared with
the conventional procedure of taking $N_F$ to be a step-function at the
quark-mass thresholds.)}
\label{fig:nfsum}
\end{figure}

Adding all flavors together gives the total
$\widetilde {N}_{F,\overline{\mbox{\tiny MS}}}^{(0)}(Q)$
which is shown in Fig.~\ref{fig:nfsum}. For reference the
continuous $N_F$ is also compared with
the conventional procedure of taking $N_F$ to be a step-function at the
quark-mass thresholds.
The figure shows clearly that there are hardly any plateaus at all
for the continuous
$\widetilde {N}_{F,\overline{\mbox{\tiny MS}}}^{(0)}(Q)$ in
between the quark masses.
Thus there is really no scale below 1 TeV where
$\widetilde {N}_{F,\overline{\mbox{\tiny MS}}}^{(0)}(Q)$
can be approximated by a constant.
In other words, for all $Q$ below 1 TeV there is always one quark
with mass $m_i$ such that $m_i^2 \ll Q^2$ or $Q^2 \gg m_i^2$ is not
true.
We also note that if one would use any other scale than the
BLM-scale for $\widetilde {N}_{F,\overline{\mbox{\tiny MS}}}^{(0)}(Q)$,
the result would be to increase the difference between the analytic
$N_F$ and the standard procedure of using the step-function at the
quark-mass thresholds.

\subsection{Comparing the Analytic $\amst(Q)$ with \ams\ }

We can obtain the renormalization group equation for the analytic
extension of the $\overline{\mbox{MS}}$ coupling $\widetilde
{\alpha}_{\overline{\mbox{\tiny MS}}}$ by using  $\widetilde
{N}_{F,\overline{\mbox{\tiny MS}}}^{(0)}(Q)$, etc.:
\begin{equation}
{d \, \widetilde{\alpha}_{\overline{\mbox{\tiny MS}}}(Q) \over d \ln Q} =
- \widetilde{\psi}_{\overline{\mbox{\tiny MS}}}^{(0)}(Q)
 {\widetilde{\alpha}_{\overline{\mbox{\tiny MS}}}^2(Q) \over \pi} + \cdots .
\label{QCDevol}
\end{equation}

The solution to Eq.~(\ref{QCDevol}) provides an analytic scale-fixed
extension of $\ams(\mu)$, which we have denoted as \amst.  The result
can be compared with the standard method of computing \ams, based
on the evolution with distinct $\psi$ functions for different quark
mass regimes. When doing this comparison one has to keep in mind that
it is possible to take quark mass threshold effects into account also
in the \ms\ scheme when calculating an observable. In the next section
we will compare our analytic extension of the \ms\ scheme with the standard
treatment of quark mass threshold effects for the hadronic width of the
Z-boson.
However, for most observables the quark mass threshold effects are not
known and thus it is also important to compare \amst\ and \ams\ directly.

\begin{figure}[htb]
\begin{center}
\mbox{\epsfig{figure=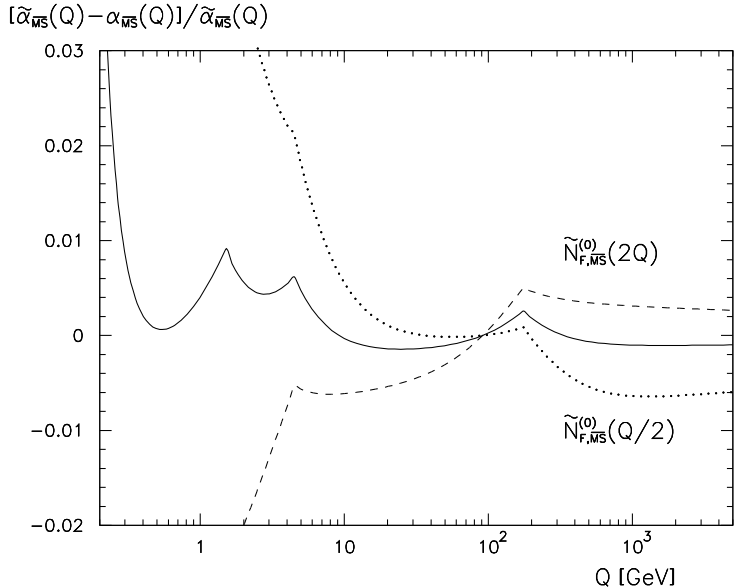,width=10cm}}
\end{center}
\caption[*]{The solid curve shows the relative difference between the
solutions to
the 1-loop renormalization group equation using continuous $N_F$,
$\widetilde{\alpha}_{\overline{\mbox{\tiny MS}}}(Q)$, and conventional discrete
theta-function thresholds, $\alpha_{\overline{\mbox{\tiny MS}}}(Q)$.
The dashed (dotted) curves shows the same quantity but using the scale $2Q$
($Q/2$)
in $\widetilde {N}_{F,\overline{\mbox{\tiny MS}}}^{(0)}$.  The solutions
have been
obtained numerically starting from the world average \protect\cite{Burrows}
$\alpha_{\overline{\mbox{\tiny MS}}}(M_Z) = 0.118$.}
\label{fig:adiff}
\end{figure}

Fig.~\ref{fig:adiff} shows the relative difference between the two
different solutions of the 1-loop renormalization group equation,
i.e. $(\widetilde{\alpha}_{\overline{\mbox{\tiny MS}}}(Q)-
           {\alpha}_{\overline{\mbox{\tiny MS}}}(Q) )/
           \widetilde{\alpha}_{\overline{\mbox{\tiny MS}}}(Q)$.
The solutions have been obtained numerically starting from the
world average \cite{Burrows}
$\alpha_{\overline{\mbox{\tiny MS}}}(M_Z) = 0.118$.
The figure shows that
taking the quark masses into account in the running leads to
effects  of the order of one percent, most especially
pronounced near thresholds.

In addition the figure shows the results obtained by using two different
scales in $\widetilde {N}_{F,\overline{\mbox{\tiny MS}}}^{(0)}(Q)$,
namely $2Q$ and $Q/2$, when solving Eq.~(\ref{QCDevol}).
This shows clearly that the BLM-scale minimizes the difference between
solutions using the continuous and discontinuous $N_F$. The other two
scale choices gives differences of several percent for small $Q$.

We see from the figure that the effect of treating thresholds
continuously can be of the order of a few percent in the magnitude of
the QCD coupling when running down from $M_Z$ to $m_c$.  This is
a significant difference, at the level of the precision of current
$\alpha_{\mbox{\scriptsize{s}}}$ determinations.  The primary factor
which influences the running is the value of the one-loop $\psi$
function, $\psi^{(0)}=\frac{11}{2}-\frac{1}{3}N_F$: a larger value of
$N_F$ gives a smaller $\psi^{(0)}$ and makes
$\alpha_{\mbox{\scriptsize{s}}}$ run more slowly; conversely a smaller
value of $N_F$ gives a larger $\psi^{(0)}$ and
$\alpha_{\mbox{\scriptsize{s}}}$ runs more quickly.

We can trace the difference in the couplings as follows: at $M_Z$, the
continuous function $\widetilde {N}_{F,\overline{\mbox{\tiny
MS}}}^{(0)}$ is above the discrete threshold value of 5, but goes below
it at 30 GeV; it remains below the discrete threshold value until
$m_b$ where it becomes larger and remains larger until $\sim$ 3 GeV,
where it becomes smaller again etc.

Thus, running down from $M_Z$, \amst\ runs slower than \ams\ until 30
GeV where the difference between them begins to close as \amst\ runs
faster than \ams ; at $\sim$ 8 GeV the difference starts to increase
again until the $b$ quark threshold where \amst\ starts to run slower
than \ams\ and the difference between the two decreases until $\sim$ 3
GeV, etc.  this behavior forms the peaks seen in Fig.~\ref{fig:adiff}.
Thus we see that \amst\ will end up higher than \ams\ when running
down to low momentum transfers starting from $M_Z$.

\section{Applications}
In this section we will show  how to compute an observable using the
analytic extension of the \ms\ scheme and compare with the standard
treatment of quark mass threshold effects in the \ms\ scheme.
The essential difference between the
perturbative expansions in the \ams\ and \amst\ couplings are terms that
contain  quark masses.
In the analytic scheme the quark mass
effects are automatically included whereas in the \ms\ scheme they have
to be included by hand for each observable.

For some observables, such as the hadronic width of the Z-boson and
the $\tau$ lepton semihadronic decay rate, corrections due to non-zero
quark masses have been calculated within the \ms\ scheme
\cite{Chetyrkin,Hoang,Soper_Surguladze,L_R_V_NPB}. To be specific we
are interested in the so called double bubble diagrams where the outer
quark loop which couples to the weak current is considered massless
and the inner quark loop is massive. Other types of mass corrections,
such as the double triangle graphs where the external current is
electroweak, are not taken into account by the analytic extension of
the \ms\ scheme. (For a recent review of higher order corrections to
the Z-boson width see \cite{Kniehl}.)

To illustrate how to compute an
observable using the analytic extension of the \ms\ scheme and compare
with the standard treatment in
the \ms\ scheme we consider the
QCD corrections to the quark part of the non-singlet hadronic width of
the Z-boson, $\Gamma_{had,q}^{NS}$. Writing the QCD corrections in terms
of an effective charge we have
\begin{equation}
\Gamma_{had,q}^{NS}=\frac{G_FM_Z^3}{2\pi\sqrt{2}}
\sum_{q}\{(g_V^{q})^2+(g_A^{q})^2\}
\left[1+\frac{3}{4}C_F\frac{\alpha_{\Gamma,q}^{NS}(s)}{\pi}\right]
\end{equation}
where the effective charge $\alpha_{\Gamma,q}^{NS}(s)$ contains all
QCD corrections,
\begin{eqnarray}
\frac{\alpha_{\Gamma,q}^{NS}(s)}{\pi} & = &
\frac{\alpha_{\overline{\mbox{\tiny MS}}}^{(N_L)}(\mu)}{\pi}
\left\{1+\frac{\alpha_{\overline{\mbox{\tiny MS}}}^{(N_L)}(\mu)}{\pi}
\left[\sum_{q=1}^{N_L}\left(-\frac{11}{12}+\frac{2}{3}\zeta_3
+ F\left(\frac{m_q^2}{s}\right)
-\frac{1}{3}\ln\left(\frac{\mu}{\sqrt{s}}\right)\right)
\right. \right. \nonumber \\ && \left. \left.
+\sum_{Q=N_L+1}^{6}G\left(\frac{m_Q^2}{s}\right)\right] + \ldots \right\}
\end{eqnarray}
The functions $F$ and $G$ are the effects of non-zero quark masses
for light and heavy quarks, respectively.
In the following we will not restrict ourselves to the case $\sqrt{s}=M_Z$
since we want to compare the two treatments of masses for arbitrary $s$.
Thereby the number of light flavors $N_L$ will also vary with $s$.
We will also assume that the matching of $N_F$ is done
at the quark masses. Thus a quark with mass $m<\mu$ is considered
as light whereas a quark with mass  $m>\mu$ is considered as heavy.

To calculate $\alpha_{\Gamma,q}^{NS}(s)$ in the
analytic extension of the \ms\ scheme one first has
to apply the BLM scale-setting procedure which absorbs all the massless
effects of non-zero $N_F$ into the running of the coupling.
This gives,
\begin{eqnarray}
\label{eq:agms}
\frac{\alpha_{\Gamma,q}^{NS}(s)}{\pi} & = &
\frac{\alpha_{\overline{\mbox{\tiny MS}}}^{(N_L)}(Q^*)}{\pi}
\left\{1+\frac{\alpha_{\overline{\mbox{\tiny MS}}}^{(N_L)}(Q^*)}{\pi}
\left[\sum_{q=1}^{N_L}F\left(\frac{m_q^2}{s}\right)
+\sum_{Q=N_L+1}^{6}G\left(\frac{m_Q^2}{s}\right)\right] + \ldots \right\}
\end{eqnarray}
where
\begin{equation}
Q^*=\exp\left[3\left(-\frac{11}{12}+\frac{2}{3}\zeta_3\right)\right]\sqrt{s}
=0.7076\sqrt{s}.
\end{equation}
Operationally, one next simply drops
all the mass dependent terms in the above expression and replaces the
fixed $N_F$ coupling $\alpha_{\overline{\mbox{\tiny MS}}}^{(N_L)}$
with the analytic \amst. (For an observable calculated with massless quarks
this step reduces to replacing the coupling.)
In this way both the massless $N_F$ contribution
as well as the mass dependent contributions from double bubble diagrams
are absorbed into the coupling
and we are left with a very simple expression,
\begin{eqnarray}
\label{eq:aganalytic}
\frac{\alpha_{\Gamma,q}^{NS}(s)}{\pi} & = &
\frac{\amst(Q^*)}{\pi}.
\end{eqnarray}
This simple expression reflects the fact that the effects of quarks in the
perturbative coefficients, both massless and massive, should be absorbed
into the running of the coupling.

To compare with the ordinary \ms\ treatment we need the functions $F$ and $G$
in Eq.~(\ref{eq:agms}).
Expansions in terms of $m^2/s$ and $s/m^2$ can be found in
\cite{Chetyrkin,Hoang,L_R_V_NPB} whereas they have been calculated numerically
in \cite{Soper_Surguladze}. In addition the
$\alpha_{\mbox{\scriptsize{s}}}^3$ correction due to
heavy quarks has been calculated as an expansion in $s/m^2$ in \cite{L_R_V_NPB}.
It should also be noted that the function $G$ was first calculated for QED
\cite{Kniehl_PLB}.
Here we will use the following expansions,
\begin{eqnarray}
F\left(\frac{m^2}{s}\right) & = &
\left(\frac{m^2}{s}\right)^2\left[\frac{13}{3}
-4\zeta_3-\ln\left(\frac{m^2}{s}\right)\right]
\nonumber \\ &&
+ \left(\frac{m^2}{s}\right)^3\left[\frac{136}{243}+\frac{16}{27}\zeta_2
+\frac{56}{81}\ln\left(\frac{m^2}{s}\right)
-\frac{8}{27}\ln^2\left(\frac{m^2}{s}\right)\right]
\\
G\left(\frac{m^2}{s}\right) & = &
\frac{s}{m^2}\left[ \frac{44}{675}
+\frac{2}{135}\ln\left(\frac{s}{m^2}\right)\right]
+\left(\frac{s}{m^2}\right)^2\left[-\frac{1303}{1058400}
-\frac{1}{2520}\ln\left(\frac{s}{m^2}\right)\right]
\end{eqnarray}
which are good to within a few percent for $m^2/s<0.25$
and $s/m^2<4$ respectively.
We will also use the relation \cite{Soper_Surguladze},
\begin{equation}
F\left(\frac{m^2}{s}\right) =
G\left(\frac{m^2}{s}\right) + \frac{1}{6}\ln\left(\frac{m^2}{s}\right)
-\left(-\frac{11}{12}+\frac{2}{3}\zeta_3 \right)
\end{equation}
to get $F$ in the interval  $0.25 < m^2/s < 1$ since the expansion of
$F$ in terms of $m^2/s$ breaks down for $m^2/s > 0.25$.

\begin{figure}[htb]
\begin{center}
\mbox{\epsfig{figure=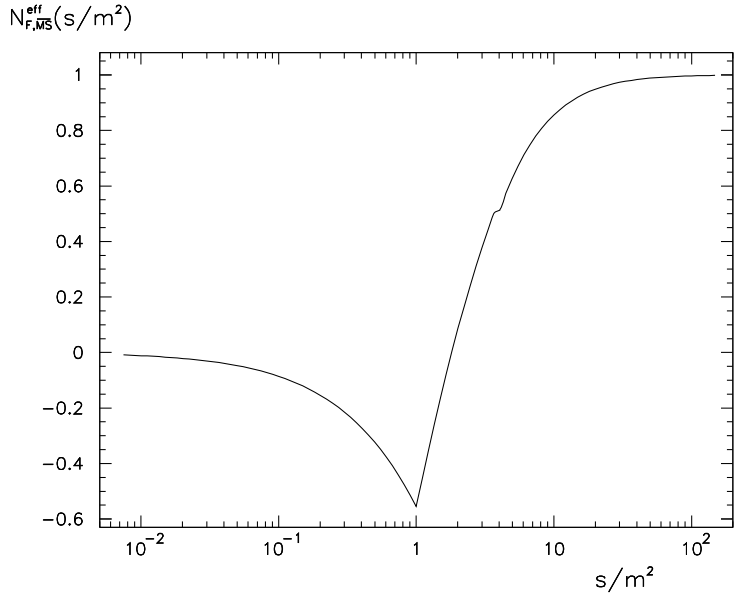,width=10cm}}
\end{center}
\caption[*]{The effective contribution to $N_F$ in the
$\alpha_{\mbox{\scriptsize{s}}}^2$-coefficient in the standard \ms\ scheme
from a quark with mass $m$ as a function of $s/m^2$ (using $\mu=\sqrt{s}$).
The discontinuity
between the two expansions in $s/m^2$ and $m^2/s$ can be seen
at the nonanalytic point $s/m^2=4$.}
\label{fig:n_eff_MS}
\end{figure}

Before carrying out the comparison of the analytic extension of the \ms\
scheme with the standard treatment it is instructive to look at the effective
contribution to  $\alpha_{\Gamma,q}^{NS}(s)$ from one flavor with mass $m$ as
a function of $s$. To make the arguments more transparent we will use
the renormalization scale $\mu=\sqrt{s}$ when doing this.
For small $s$, when the quark is considered heavy the
contribution is given by $G(m^2/s)$ whereas for larger $s$ the quark is
considered as light  and contributes with $F(m^2/s)-
\frac{11}{12}+\frac{2}{3}\zeta_3$.
Normalizing to the massless contribution  $-\frac{11}{12}+\frac{2}{3}\zeta_3 $
gives the contribution to the effective $N_F$ in the
$\alpha_{\mbox{\scriptsize{s}}}^2$-coefficient,
\begin{equation}
N_{F,\overline{\mbox{\tiny MS}}}^{\mbox{\scriptsize{eff}}}
\left( \frac{s}{m^2} \right) =
\left\{
\begin{array}{lc}
{\displaystyle \frac{G\left({ \frac{m^2}{s}}\right)}
{ -{ \frac{11}{12}+\frac{2}{3}\zeta_3}} }
&  \quad \text{for $\sqrt{s}<m$  } \\ \\
{\displaystyle
\frac{F\left({ \frac{m^2}{s}}\right)-{ \frac{11}{12}+\frac{2}{3}\zeta_3}}
{ -{ \frac{11}{12}+\frac{2}{3}\zeta_3}}}
& \quad \text{for  $\sqrt{s}>m$ }
\end{array}
\right. \quad ,
\end{equation}
which is shown in Fig.~\ref{fig:n_eff_MS} as a function of $s/m^2$.

At first it might seem unnatural that the effective contribution to $N_F$
in the $\alpha_{\mbox{\scriptsize{s}}}^2$-coefficient is
negative for heavy quarks. However, this is a characteristic feature
of the standard \ms\ scheme which arises from the fact that the number of
flavors in the running of the coupling is kept constant.
Starting from a scale well below the threshold the number of flavors in the
running as well as in the $\alpha_{\mbox{\scriptsize{s}}}^2$
coefficient is not affected by the heavy quark.
As the threshold is approached from below the number of flavors in the running
should increase which would make the running of the coupling slower (since
$\psi^{(0)}$ would be smaller) which in turn should lead to a larger \ams.
But, since the number of flavors is kept constant in the running this
effect has to be taken into account by adding a positive contribution
to the $\alpha_{\mbox{\scriptsize{s}}}^2$-coefficient,
i.e. the function $G$. Since the massless contribution
is negative this means that the
contribution to $N_F$ becomes negative for a heavy
quark. Once the threshold has been
crossed the number of flavors in the running changes and the need to compensate
for a too small \ams\ vanishes rapidly as the scale is increased above the
threshold. For scales well above the threshold the mass-effects are negligible
and the massless result is regained as $F$ goes to zero. This should be
compared with the analytic \ms\ scheme where $N_F$ is increased continuously
in the running.

To compare the analytic extension  of the \ms\ scheme
with the standard \ms\ result for
$\alpha_{\Gamma,q}^{NS}(s)$ we will apply
the BLM scale-setting procedure also for the standard \ms\ scheme.
This is to ensure that any differences are due to the different
ways of treating quark masses and not due to the scale choice.
In other words we want to compare Eqs.~(\ref{eq:agms})
and (\ref{eq:aganalytic}). As the normalization point we use
$\alpha_{\overline{\mbox{\tiny MS}}}^{(5)}(M_Z)=0.118$
which we evolve down to $Q^*=0.7076M_Z$ using leading order
massless evolution with $N_F=5$. This value is then used to calculate
$\alpha_{\Gamma,q}^{NS}(M_Z)=0.1243$ in the \ms\ scheme using
Eq.~(\ref{eq:agms}). Finally, Eq.~(\ref{eq:aganalytic}) gives the
normalization point for $\amst(Q^*)$.

\begin{figure}[htb]
\begin{center}
\mbox{\epsfig{figure=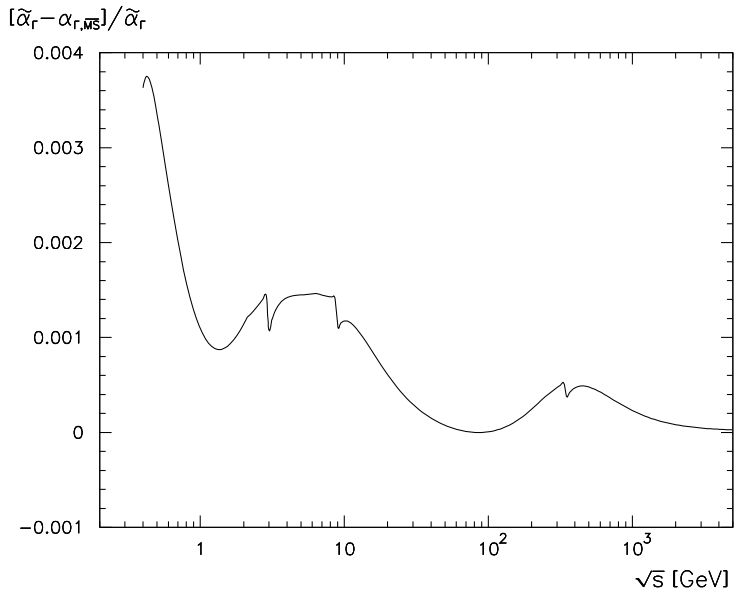,width=10cm}}
\end{center}
\caption[*]{The relative difference between the calculation of
$\alpha_{\Gamma,q}^{NS}(s)$ in the analytic extension of the \ms\ scheme
and the standard treatment of masses in the \ms\ scheme. The discontinuities
are due to the mismatch between the $s/m^2$ and $m^2/s$ expansions of the
functions $F$ and $G$.}
\label{fig:gdiff}
\end{figure}

Fig.~\ref{fig:gdiff} shows the relative difference between the two
expressions for $\alpha_{\Gamma,q}^{NS}(s)$ given by Eqs.~(\ref{eq:agms})
and (\ref{eq:aganalytic}) respectively. As can be seen from the figure the
relative difference is smaller than 0.2\% for scales above 1 GeV. Thus
the analytic extension of the \ms\ scheme takes the mass corrections
into account in a very simple way without having to include an infinite
series of higher dimension operators or doing complicated multi-loop
diagrams with explicit masses.

\section{Conclusion}

An essential feature of the \av(Q) scheme is that there is no
renormalization scale ambiguity, since  $Q^2$ is the physical momentum
transfer. The \av\ scheme naturally takes into account quark mass
thresholds, which is of particular phenomenological importance to QCD
applications in the intervening mass region between those thresholds.
In this paper we have utilized commensurate scale relations to provide
an analytic extension of the conventional \ms\  scheme in which many of
the advantages of the \av\ scheme are inherited by the \amst\ scheme,
but only minimal changes have to be made to the standard \ams\ scheme.
Given the commensurate scale relation, Eq.~(\ref{eq:csrmsofv}),
connecting \amst\ to \av\, expansions in
\amst\ are effectively expansions in \av\ to the given order in perturbation
theory at a corresponding commensurate scale.
Unlike the conventional \ams\ scheme, the modified \amst\ scheme is
analytic at quark mass thresholds, and  it thus provides a natural
expansion parameter for perturbative representations of observables.
In the
Abelian limit $N_C \to 0$, \amst\ scheme agrees with the standard effective
charge method of QED.

We have found that taking finite quark mass effects into account analytically
in the running, rather than using a fixed $N_F$ between thresholds, leads to
effects of the order of one percent for the one-loop running coupling, with the
largest differences occurring near thresholds. These differences are important
for observables that are calculated neglecting quark masses and could in
principle  turn out to be significant in comparing  low and high energy
measurements of the strong coupling.

We have also found that our extension of the \ms\ scheme, including quark mass
effects analytically, reproduces the standard treatment of  quark masses in the
\ms\ scheme to within a fraction of a percent. The standard treatment amounts
to either calculating multi-loop diagrams  with explicit quark masses or adding
higher dimension operators to the effective Lagrangian. These corrections
can be viewed as compensating for the fact that the number of flavors in the
running is kept constant between mass thresholds. By utilizing the BLM scale
setting procedure, based on the massless $N_F$ contribution, the analytic
extension  of the \ms\ scheme correctly absorbs both massless and mass dependent
quark contributions from QCD diagrams, such as the double bubble diagram, into
the running of the coupling. This gives the opportunity to convert
a calculation made in the \ms\ scheme with massless quarks into an
expression which includes quark mass corrections from QCD diagrams
by using the BLM scale and replacing \ams\ with \amst.

For simplicity we have analyzed the mass corrections arising from
analyticity only to leading order in QCD. For further precision, our
analysis will need to be systematically improved. For example, at
higher orders the commensurate scale relation connecting \av\ to \ams\
will have to be corrected with finite mass effects. We have seen that the
BLM scale minimizes the difference between the analytic and the
conventional \ams-coupling. Thus, these kinds of corrections are not likely
to decrease the difference between the analytic and the
conventional \ams-coupling.

Finally, we note the potential importance of utilizing the  \av\
effective charge or the equivalent analytic \amst\ scheme in
supersymmetric and grand unified theories, particularly since the
unification of couplings and masses would be expected to occur in terms
of physical quantities rather than parameters defined by theoretical
convention.

\section*{Acknowledgments}

We would like to thank
A.~Hebecker,
C.~Lee,
H.J.~Lu,
G.~Mirabelli,
M.~Peter,
D.~Pierce,
O.~Puzyrko,
A.~Rajaraman,  and
W.~Kai Wong,
for useful discussions and comments.

\end{document}